\documentclass[submission,copyright,creativecommons]{eptcs}
\providecommand{\event}{FMSPLE 2015} 
\usepackage{breakurl}             

\usepackage[utf8]{inputenc}
\usepackage{wrapfig} 
\usepackage{graphicx}
\usepackage{amsthm}
\usepackage{amsmath}

\usepackage{tikz}
\usetikzlibrary{arrows,shapes,trees}
\usepackage{caption}
\usepackage{subcaption}

\theoremstyle{definition}

\newcommand{\focal}{{F}o{C}a{L}}
\newcommand{\focalize}{\focal{}i{Z}e}

\title{Towards correct-by-construction product variants of a software product line: \\
GFML, a formal language for feature modules} 
\author{Thi-Kim-Zung Pham
\institute{Cedric, CNAM}
\institute{Paris, France}
\email{pham\_t43@auditeur.cnam.fr}
\and
Catherine Dubois
\institute{Cedric, ENSIIE}
\institute{Évry, France}
\email{catherine.dubois@ensiie.fr}
\and
Nicole Levy 
\institute{Cedric, CNAM}
\institute{Paris, France}
\email{nicole.levy@cnam.fr }
}
\def\titlerunning{A formal language for feature modules}
\def\authorrunning{T.K.Z Pham, C. Dubois \& N. Levy}
\begin{document}
\maketitle

\begin{abstract}
Software Product Line Engineering (SPLE) is a software engineering paradigm that focuses on reuse and variability. Although feature-oriented programming (FOP) can implement software product line efficiently, we still need a method to generate and prove correctness of all product variants more efficiently and automatically. In this context, we propose to manipulate feature modules which contain three kinds of artifacts: specification, code and correctness proof. We depict a methodology and a platform that help the user to automatically produce correct-by-construction product variants from the related feature modules. As a first step of this  project, we begin by proposing a language, GFML, allowing the developer to write such feature modules. This language is designed so that the artifacts can be easily reused and composed. GFML files contain the different artifacts mentioned above. The idea is to compile them into \focalize, a language for specification, implementation and formal proof with some object-oriented flavor. In this paper, we define and illustrate this language. We also introduce a way to compose the feature modules on some examples.  
\end{abstract}

\section{Introduction}Software Product Line Engineering (SPLE) is a paradigm used to develop software-intensive systems that share common assets \cite{pohl_software_2005, apel_feature-oriented_2013}. In SPLE, a feature is used to represent a characteristic behavior as a unit of functionality of the software product line (SPL) \cite{apel_feature-oriented_2013}. Given a set of features, the configuration of a SPL is constructed by composing the features. Following generative programming mechanisms \cite{barth_generative_2001}, the respective product variant can be derived automatically from a feature selection and  artifacts of each selected feature. The product generation may be realized through Feature-Oriented Programming (FOP) techniques \cite{batory_scaling_2003}, in which each feature is mapped to a feature module containing its artifacts. The product variant is synthesized from the artifacts of the involved feature modules. In this context, we demonstrate our approach that helps the developer to write feature modules containing their artifacts using a dedicated and generic language. 

Some authors proposed  approaches for constructing product variants of a product line by reusing and synthesizing artifacts. Using design by contract \cite{meyer_applying_1992} and adhering to FOP techniques, Th\"um proposed a proof composition approach and strategies to reduce the effort of verification  by reusing partial proofs \cite{thum_proof_2011} \cite{thum_family-based_2012}  \cite{thum_classification_2014}. The FEATUREHOUSE framework, described in \cite{apel_featurehouse:_2009}, enables the construction of a new program from existing programs using the structure of the SPL feature tree. Although the above methods can implement software product line efficiently and mostly automatically, we still need a method to generate and prove correctness of all product variants more efficiently and automatically. 

In this direction some advances have been proposed in the context of programming language meta-theory within the Coq proof assistant \cite{delaware_product_2011}  \cite{delaware_modular_2013}.
However these tools  are dedicated to a very specific domain and still require an important expertise. 

To tackle the above limitations, we depict a methodology that helps the user to automatically produce correct-by-construction product variants. This methodology aims at being independent of any concrete target language but at first we focus and experiment with 
\focalize, a language used to specify, implement and prove (\url{http://focalize.inria.fr}). In this paper, we propose a language, called \textbf{GFML} (for Generic Feature Module Language) allowing the developer to write feature modules which contain three kinds of artifacts:  specifications, code and correctness proofs. GFML is inspired from \focalize{} but sticks to a FOP approach and tries to reduce the developer's effort. This language is designed so that the artifacts can be easily reused, composed and translated into different languages. 

In Section \ref{background}, we first describe the background of our paper, in particular we present  briefly \focalize. Then in Section \ref{Artifacts}, we illustrate our definition of feature module and a brief description of our language (GFML) used to write feature modules. In Section \ref{sec:GFMLtranslation}, we illustrate the translation of GFML into \focalize. In Section \ref{sec:composition}, we describe how to compose two feature modules by giving an example, and  depict a composition framework for automatically generating correct-by-construction product variants. In Section \ref{sec:Relate}, we review related work. Finally we conclude the paper and discuss future works in Section \ref{sec:conclusion}.

\section{Background}\label{background}
\subsection{Software Product Line}
\begin{wrapfigure}{r}{0.35\textwidth}
  \vspace{-15pt}
  \begin{center}
    \includegraphics[width=0.35\textwidth]{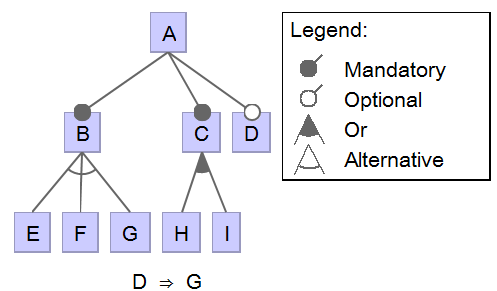}
  \end{center}
  \vspace{-15pt}
  \caption{Feature diagram}
  \label{fig:ex-dia}
  \vspace{-10pt}
\end{wrapfigure}

In SPLE, a \textbf{feature} is a characteristic behavior specified as a unit of functionality of a product line \cite{apel_feature-oriented_2013}. A set of features, called a \textbf{feature model}, is often graphically depicted as a tree, also called a \textbf{feature diagram}. In addition to the presentation of commonality and variability, a feature diagram also contains various relationships and additional constraints between  features. The main relationships in feature diagrams are optional/mandatory, alternative/or and implies/excludes. Because of visual aid, feature diagrams are considered as standard representations that help the user to choose product configurations of a product line. In this work, we only focus on a simple form of feature diagrams (i.e. Figure \ref{fig:ex-dia}) that can illustrate optional features. 

\textbf{Valid Configuration.} A configuration of a product line is a set of features selected in the corresponding feature diagram. All the information such as relations and constraints inferred from the feature diagram is used to check the validity of the configuration. In our work, we only consider valid configurations. However, checking the validity of a configuration is out of the scope of this paper. Many checkers exist (e.g. \cite{batory_feature_2005}, \cite{BenavidesT13}) and we rely on them. As we start from valid configurations, we consider that the configurations contain all the features that are to be composed in order to build the final products. So the distinction between mandatory and optional features is not relevant here. 

\subsection{Motivating Example}
\begin{wrapfigure}{r}{0.35\textwidth}
  \vspace{-40pt}
  \begin{center}
    \includegraphics[width=0.35\textwidth, height=0.16\textwidth]{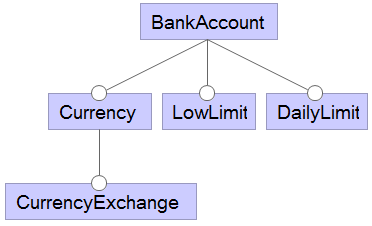}
  \end{center}
  \vspace{-15pt}
  \caption{Feature diagram of bank account product line}
  \vspace{-50pt}
  \label{fig:BA-dia}
\end{wrapfigure}

As a running example, we consider the bank account product line described in \cite{thum_family-based_2012} whose feature diagram is shown in Figure \ref{fig:BA-dia}. It illustrates a family of products allowing the management of bank accounts. The root feature \textit{BankAccount} (BA for short) provides the basic management of an account. It allows the bank storing the current balance and the amount of money added into or withdrawn from the account. A customer can withdraw more money from the account than available if it is within an over limit. The feature BA has three optional child features \textit{DailyLimit} (DL for short), \textit{LowLimit} (LL for short), \textit{Currency}. 
The feature DL allows the bank to limit the amount of money withdrawn in a day while the other feature LL indicates that the bank only authorizes a customer to withdraw money from the account only if the amount is greater than a low limit. Related
The feature \textit{Currency} accommodates the management of currency. Finally, an optional feature \textit{CurrencyExchange} is established as a child of \textit{Currency} to enable the calculation of currency exchange.

\subsection{Quick Presentation of \focalize}\label{focalize}

This subsection presents briefly the technical background necessary to understand the section about the translation of GFML to \focalize.

The \focalize{} (\url{http://focalize.inria.fr}) environment provides a set of tools to describe and implement
functions and logical statements together with their proof. A \focalize{}
source program is analyzed and translated into OCaml sources for execution and
Coq  sources for certification. The \focalize{} language has an object oriented
flavor allowing inheritance, late binding and redefinition. These characteristics are very helpful to reuse specifications, implementations and proofs. 

A \focalize{} specification can be seen as  a set of algebraic properties describing relations between input and output of the functions implemented in a \focalize{} program. 
For writing code,  \focalize{} offers a pure functional programming style close to ML, featuring strong typing, recursive functions, data types and pattern-matching.
Proofs written using the \focalize{} proof language
are sent to the Zenon prover which produces proofs that can be verified by Coq for more confidence \cite{bonichon_zenon_2007}. The \focalize{}
proof language is a declarative language in which the programmer states a
property and gives hints  to achieve its proof which is performed by Zenon.

\focalize{} units are called \emph{collections}. They  contain
 entities in a model akin to classes and objects or types and values. Collections have
 functions and  properties which can be called using the ``\texttt{!}''  notation. They are derived from other units called \emph{species} which specify and implement functions.

A species defines a set of entities together with functions and
properties applying to them. At the beginning of a development,  the
representation  of these entities is usually  abstract, it is
precised later in the development. However the type of these entities
is referred as \texttt{Self} in any species. Species may contain specifications, functions and proofs.
More precisely species may specify a function or a property (with resp. \texttt{signature},
\texttt{property} keywords) or implement them 
(\texttt{let} keyword when a function is defined, \texttt{proof of} keyword to introduce a proof of a property).
A \texttt{let} defined function must match its signature and similarly a proof
introduced by \texttt{proof of} should prove the statement given by the
\texttt{property} keyword.  Statements belong to first order typed logic.

As said previously, \focalize{} integrates inheritance, late binding and redefinition to ease reuse and modularity. Inheritance allows the definition of a new species from one or several other species. The new species inherits all the functions, properties and proofs of its parents. Some syntactical mechanisms  are provided to prevent ambiguities. A species may provide a definition for a function that is only specified in its parents. It may also redefine a function when this one is already defined in a parent but in that case the signature is maintained (no overloading). Multiple inheritance comes with a late binding mechanism close to the one found in object oriented languages (even if the resolution is statically  done by the compiler).

A collection is a species where every specified function is defined and every property is proved (or explicitly admitted). Furthermore, in a collection, the concrete \emph{representation} of entities is made private and a programmer using the collection can only use its functions and properties according to the interface. Consequently building a collection from a species provides encapsulation.  
Species may have parameters which may either be collections or entities providing
in that way parametric polymorphism.  This parametrization will be intensively used in the translation of GFML to \focalize{} (see Section~\ref{sec:GFMLtranslation}).

For more details on \focalize{} please refer to the reference
manual. A more thorough overview  can be found in
\cite{hardin2008}.

\section{Artifacts}\label{Artifacts}
\subsection{Feature Module}
We adhere to the FOP technique that suggests to map each feature to a separate \textbf{feature module} that implements the feature. In our context, each feature module consists of its \textbf{artifacts}: specifications, code and correctness proofs. Specifications are given as a set of expected properties or requirements. Technically these properties are logical formulas relating together some functions described only by their signature. Thus in our setting, a specification is close to an algebraic data-type. Code has to be understood here  as the implementation of the functions introduced/declared in the specifications. We place ourselves in a functional programming setting. The proofs here concern the correctness of the code with respect to the specifications. 

\begin{itemize}
\item \textbf{Specification artifact} includes the function declarations  and the properties. A function declaration or signature only describes the name of the function  and the type of its arguments and result. A property is a (first order typed) logical formula. A new property is expressed by a new logical formula while a refining property refers to  existing properties of the $parent$ feature module and may add new premises and conclusions.
\item \textbf{Code artifact} consists of the definition of the concrete representation type $r$  and the function definition/redefinitions $df$. The representation type is the concrete type associated to the abstract data-type specified in the specification artifact. It will be a concrete type (basic or complex types, \textit{à la ML}) or a Cartesian product of concrete types. The representation type is  unique for each feature module and can be also constructed from the existing representation type of $parent$. A function is defined/redefined after the representation type has been defined. 
\item \textbf{Proof artifact} contains proofs $pf$. Each proof corresponds to  a property. While writing the proof for a refining property, the mentioned properties of $parent$ can be reused to support the proving process.
\end{itemize}

We define a feature module $fm$ as a 5-tuple: function declarations $d$, properties $p$, representation type $r$, function definitions/redefinitions $df$ and proofs $pf$.
\begin{center}
$fm = (d, p, r, df, pf)$
\end{center}

\subsection{Description of GFML}

















We propose a generic language, called \textbf{GFML}, to write these feature modules. Each feature module is embedded into a separate file \textit{.gfm}. This language is suitable for all feature modules possibly containing three kinds of artifacts: specifications, code and correctness proofs. 

This language is inspired from \focalize{}, in particular the styles for writing specifications, code and proofs are common. GFML and  \focalize{} mainly differ in the way to structure and organize information. Our main objective in defining GFML is to propose a  language close to \focalize{} that already allows for the three kinds of artifacts in a single setting but closer to the description of commonality and variability we can find in feature diagrams. As we will see in Section~\ref{sec:GFMLtranslation}, the expression of a feature module is much simpler that its translation in \focalize. 
GFML is also inspired from design by contract applied to FOP as in
\cite{ThumSKAS12}. In the same way, GFML syntax allows the programmer to focus in a GFML feature module on the modifications brought to the specifications or implementations of the parent.

\subsection{Examples}
Corresponding to the feature diagram of the bank account product line given in Figure \ref{fig:BA-dia}, the feature module BA implements the root feature BA. Feature modules DL, LL, and \textit{Currency} 
are mapped from the features DL, LL, and \textit{Currency} 
respectively. They have a common parent, i.e. the root module BA. Feature module \textit{CurrencyExchange} corresponding to feature \textit{CurrencyExchange}, is a child of the module \textit{Currency}. 

\begin{figure}
        \begin{subfigure}[b]{0.475\textwidth}
          \begin{center}
                \fbox{\includegraphics[width=1\textwidth]{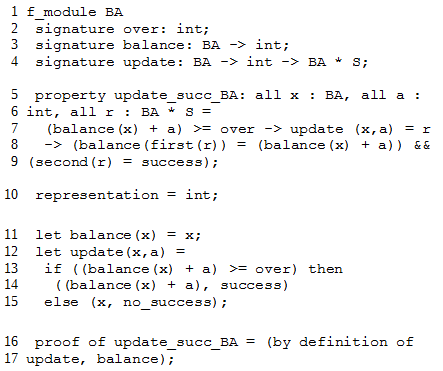}}
                \caption{Module BA}
                \label{fig:BA}
                \end{center}
        \end{subfigure}\qquad
        \begin{subfigure}[b]{0.475\textwidth}
                \fbox{\includegraphics[width=1\textwidth]{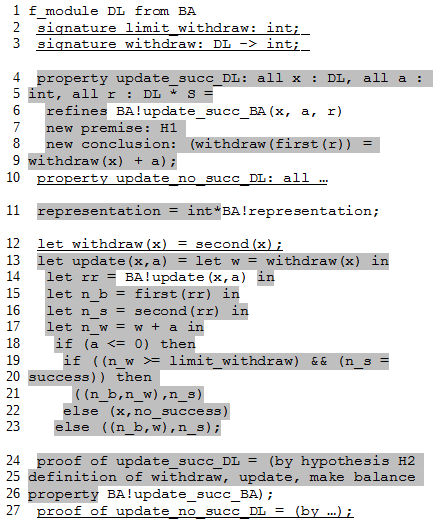}}
                \caption{Module DL}
                \label{fig:DL}
        \end{subfigure}
         \vspace{-20pt}
        \caption{Feature modules in GFML}\label{fig:BA+DL}
          \vspace{-10pt}
\end{figure}

The root feature module BA, written with GFML, is shown in Figure \ref{fig:BA}. This module includes three signatures: $over$ - is over limit, $balance$ - gets the current balance value  of the account and $update$ - upgrades the new value balance and also returns the status of the operation (\textit{success} or \textit{no\_success}, of type $S$ whose definition is omitted here). The next part of the module BA contains the property $update\_succ\_BA$ (line 5) that specifies a customer can withdraw more money $a$ from the account than available if the balance is within $over$. In that case, its status must be $success$. 
The primitive functions $first$ and $second$ used in this property are the usual projections of a Cartesian product.
Then the representation type is defined as $int$ (line 10), it means that an account is only represented by its balance. Then appear the definitions of the functions $balance$ and $update$ (lines 11-15). The proof of property $update\_succ\_BA$ includes two proof hints: by definition of $update$ and $balance$ definitions. This means that the proof must be done by unfolding the definitions of both functions.

Another example of GFML feature module is the module DL defined according to its $parent$ feature module BA (Figure \ref{fig:DL}). Two new declarations $limit\_withdraw$ and  $withdraw$ are added into the module (lines 2-3).  The module introduces the constant $limit\_withdraw$ only declared at that point.  It denotes the limit of withdrawn money in a day. It also introduces  another function  $withdraw$  that returns, for an account, the current amount of withdrawn money in a day. The functions $update$, $balance$ and $over$ defined in the parent are also available in the present feature module. A new property $update\_succ\_DL$ is obtained by modifying the property $update\_succ\_BA$ from the $parent$ (line 6). This modification includes a new premise called H1  (line 7) for short in the figure (but announced below)  and a new conclusion (lines 8-9). The premise H1 is expressed as follows:
\begin{align*}
&r = update(x,a) \rightarrow (a <= 0) \rightarrow
(all \enspace n\_w \enspace w : int, all \enspace n\_s : S, \enspace withdraw (x) = w \enspace \&\&\enspace w + a = \\
&n\_w \enspace \&\& \enspace second(BA!update(x,a)) = n\_s \rightarrow (n\_w >= limit\_withdraw) \enspace \&\&  \enspace (n\_s = success))
\end{align*}
It states that the bank allows a customer to withdraw money only if the amount of withdrawn money in a day is greater than $limit\_withdraw$ (in this case \textit{w}, \textit{n\_w} and $limit\_withdraw$ are negative numbers). The new conclusion states that this operation has to modify the account by updating the amount of withdrawn money. The representation type of module DL is defined as a Cartesian product of $int$ (i.e. the concrete type associated with the amount of money withdrawn in a day) and the representation type of the \textit{parent} (line 11). The functions $withdraw$ and $update$ are defined/redefined (lines 12-23). We can notice that in the redefinition of $update$ the call to the parent function $update$ is done with the parameter $x$, there is here an implicit conversion that will be inserted during the translation into \focalize.
The proof of the property $update\_succ\_DL$ reuses the property $update\_succ\_BA$ as a proof hint (line 26). All modifications and additions compared to the parent module BA are highlighted or underlined. 

\section{From GFML to \focalize}\label{sec:GFMLtranslation}
\begin{figure}
        \begin{subfigure}[b]{0.475\textwidth}
          \begin{center}
                \fbox{\includegraphics[width=1\textwidth]{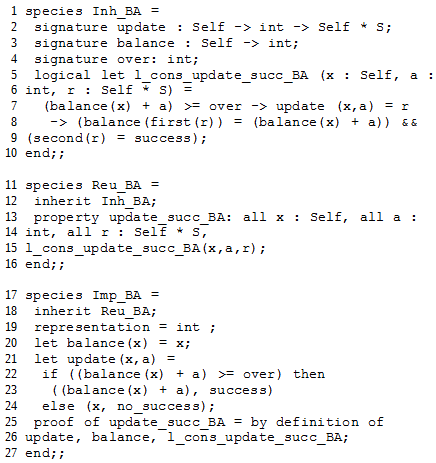}}
                \caption{Module BA in FoCaLize}
                \label{fig:BA_foc}
                \end{center}
        \end{subfigure}\qquad
        \begin{subfigure}[b]{0.475\textwidth}
                \fbox{\includegraphics[width=1\textwidth]{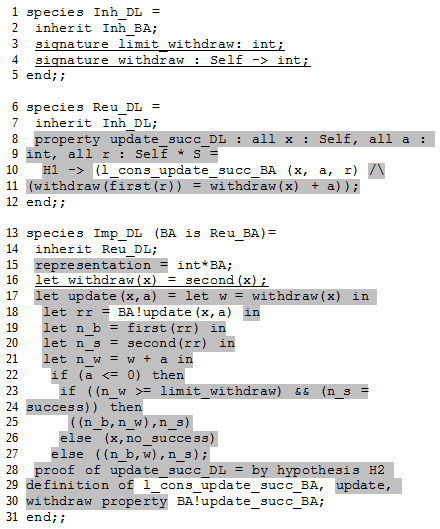}}
                \caption{Module DL in FoCaLize}
                \label{fig:DL_foc}
        \end{subfigure}
         \vspace{-20pt}
        \caption{Feature modules in FoCaLize}\label{fig:BA+DL_foc}
          \vspace{-10pt}
\end{figure}

In this section, we  illustrate the translation 
from GFML to \focalize. A GFML feature module is mapped into three \focalize{} separate \textit{species}. The first \textit{species} \textit{\textbf{Inh\_}} contains all desired elements that are inherited in the child module. These elements can be function declarations, properties or predicates. A predicate, which is introduced with the keyword `$logical\ let$' and named $l\_cons$, is associated with each property $p$ found in a module. 
For example, the feature module BA written in Figure \ref{fig:BA_foc} is mapped to the species $Inh\_BA$ (line 1). A predicate $l\_cons\_update\_succ\_BA$ (line 5) is associated with the property $update\_succ\_BA$. In Figure \ref{fig:DL_foc}, the species $Inh\_DL$ of the child DL inherits the species $Inh\_BA$ of module BA (line 2). 

The \focalize{} inheritance mechanism allows the programmer  to reuse the artifacts without any change (except for functions that can be redefined without changing its signature). But although some properties may be kept in the child module, others may be modified by adding new premises or/and new conclusions. In this last case, the \focalize{} inheritance mechanism is not sufficient. We have to use the parametrization facility to encode the right meaning. To tackle this limitation, we propose the second \textit{species} \textbf{\textit{Reu\_}} that inherits the first \textit{species} \textit{\textbf{Inh\_}}. It includes all desired specifications that are reused to describe new ones in the child. Each property specified in this \textit{species} must have a corresponding predicate  $l\_cons$ in the first \textit{species} \textit{\textbf{Inh\_}}. Instead of mentioning the property, we use this predicate for expressing new properties in the child. This trick is used because \focalize{} does not  allow the modification of an inherited  property in a \textit{species}. For example, in Figure \ref{fig:BA_foc} \textit{species} $Reu\_BA$ inherits the \textit{species} $Inh\_BA$. It contains the property $update\_succ\_BA$ (line 13) that is related to the predicate $l\_cons\_update\_succ\_BA$ in \textit{species} $Inh\_BA$. This predicate is reused to specify the refining property $update\_succ\_DL$ of module DL (line 10 of Figure \ref{fig:DL_foc}). 

 The parametrization mechanism in FoCaLize
is used here to circumvent the fact that in \focalize{} when the representation type is fixed in a species P, it cannot be changed in any species which inherits  P. So in order to build a new species that can reuse functions, properties, predicates and proofs that are defined for BA, we have to parametrize this species by an implementation of BA.  Thus,  when the DL feature is selected, the account is implemented as a pair containing a basic account of type BA and an integer corresponding to the new attribute, i.e. the amount of money withdrawn in a day. The implementation of the functions $balance$ and $over$ are, in this context, just a call to the parent's functions combined with the first projection. For example, \textit{species} $Imp\_BA$ inherits \textit{species} $Reu\_BA$ (line 18 of Figure \ref{fig:BA_foc}) while  \textit{species} $Imp\_DL$ inherits \textit{species} $Reu\_DL$ (line 14 of Figure \ref{fig:DL_foc}). A parameter $BA$ encapsulates \textit{species} $Reu\_BA$ (line 13 of Figure \ref{fig:DL_foc}). It is used to call function $update$ of module BA (line 18). 
In the proof of the refining property $update\_succ\_DL$, property $update\_succ\_BA$ of the parent BA is considered as a proof hint (line 30). Its associated formula $l\_cons\_update\_succ\_BA$ is also a proof hint (line 29). Let us notice that Zenon, the \focalize{} prover has done all the proofs with the given hints. 

We can notice that the use of a language such as \focalize{} for implementing and verifying feature modules is quite complex. The proposed language GFML is a solution to write the feature modules together with their artifacts more easily.
We have followed with success this translation scheme for all the features appearing on the feature diagram of Figure \ref{fig:BA-dia}. For the moment, the translation is done manually.

\section{Towards Feature Module Composition}\label{sec:composition}
\subsection{Methodology for composing artifacts}
\begin{figure}
         \centering
         \includegraphics[height=4.3cm, width=16cm]{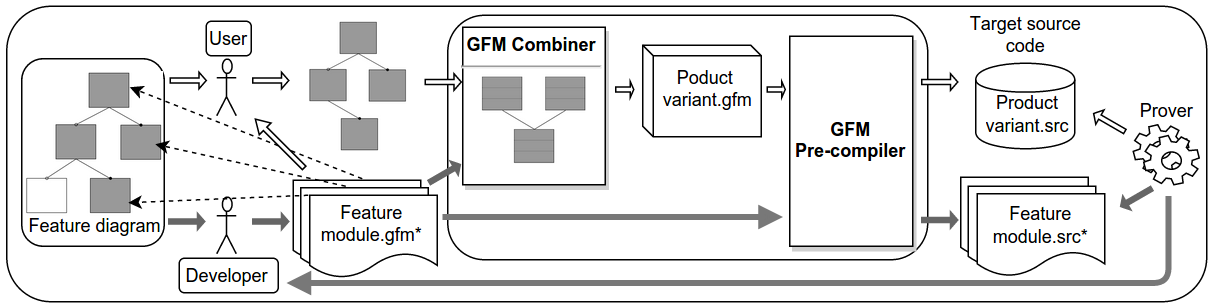}
         \vspace{-20pt}
         \caption{Methodology}
         \vspace{-10pt}
         \label{fig:met}
      \end{figure}

The work previously described is the first step in the proposition of a methodology to 
help the user to automatically produce correct-by-construction product variants from features selected in a feature diagram.  This methodology is illustrated in Figure \ref{fig:met}. We assume a SPL is described with a feature diagram. First,  the developer writes the different feature modules with the GFML language. He uses the \textbf{GFM Pre-compiler} to translate his/her feature modules into \focalize{} to verify them (using the Zenon prover). Once this work has been done, the user - he may be different from the developer - chooses  features from the feature diagram. Based on this selection, the corresponding configuration is determined. The related GFML feature modules are then sent to the \textbf{GFM Combiner} that will compose the feature modules of the configuration in order to obtain a composition feature module that will be translated into the desired product variant, which by composition is correct by construction. We expect this method to be independent from the target language which is required to be able to express specifications and code and also proofs (unless these ones can be automatically produced). 

In the next subsection we explain the main ideas of our composition mechanism on the running example. 

\subsection{Composition of two feature modules: an example} 
Some problems may appear when composing two feature modules, in particular  conflicts may appear when synthesizing their artifacts. For example, which properties should be performed first? Which   synthesized representation type will make less complex writing? To solve these problems, we suggest the user gives a composition order of the modules. In our work, we only consider the modified elements such as refining properties, representation types, function redefinitions or proofs of refining properties. 

\begin{figure}
        \begin{subfigure}[b]{0.475\textwidth}
          \begin{center}
                \fbox{\includegraphics[width=1\textwidth]{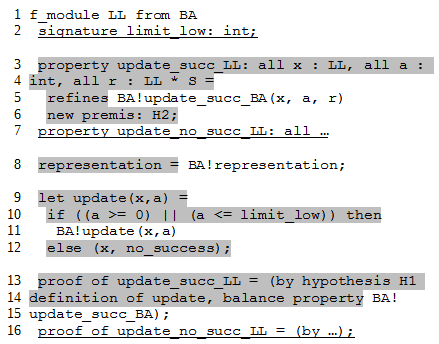}}
                \caption{Module LL}
                \label{fig:LL}
                \end{center}
        \end{subfigure}\qquad
        \begin{subfigure}[b]{0.475\textwidth}
                \fbox{\includegraphics[width=1\textwidth]{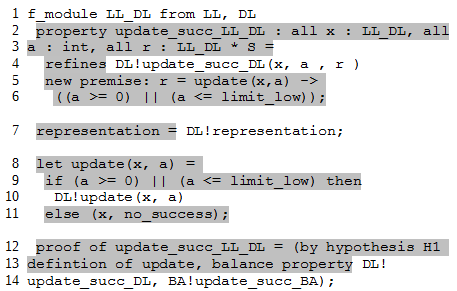}}
                \caption{Module LL\_DL}
                \label{fig:LL_DL}
        \end{subfigure}
         \vspace{-20pt}
        \caption{Composition of feature modules}\label{fig:com}
          \vspace{-10pt}
\end{figure}

Similar to module DL, module LL written in Figure \ref{fig:LL} is extended from the root module BA. A new signature $limit\_low$ declares the low limit of the account. A new property $update\_no\_succ\_LL$ is specified in line 7 and its proof is indicated in line 16. However, we only consider the modifications that can cause conflicts when composing two modules. For example, let us compose  module LL with module DL. Their composition is the module LL\_DL given in Figure \ref{fig:LL_DL}. Property $update\_succ\_LL$ of module LL includes two parts (lines 3-6 of Figure \ref{fig:LL}). Its highlighted part includes a new premise (shortened by H2), expressed as follows: 
\begin{align*}
r = update(x,a) \rightarrow
	((a >= 0) \enspace || \enspace (a <= limit\_low))
\end{align*}
The premise specifies that a customer can withdraw money $a$ from the account only if $a$ is less than $limit\_low$ (in this case \textit{a} and $limit\_low$ are negative numbers). The property $update\_succ\_BA$ from the $parent$ BA is its remaining part. Following the composition order, the composite property of two properties $update\_succ\_LL$ and $update\_succ\_DL$ is synthesized by taking H2 first and then mixing with $update\_succ\_DL$. It is named $update\_succ\_LL\_DL$ and  embedded into a composition feature module LL\_DL (lines 2-6 of Figure \ref{fig:LL_DL}). Similarly, the composite representation type is synthesized in line 7. The function $update$ is redefined (lines 8-11). The proof of the composite property $update\_succ\_LL\_DL$ reuses two properties $update\_succ\_DL$ and $update\_succ\_BA$ as its proof hints (lines 13-14) indicate. The highlighted parts of module LL (Figure \ref{fig:LL}) are the parts highlighted in module LL\_DL (Figure \ref{fig:LL_DL}). The other parts which are not highlighted in Figure \ref{fig:LL_DL} are taken from module DL.

\section{Related Work} \label{sec:Relate}
Recently, many authors focused on constructing and verifying program variants in SPL. Together with these, composition methods of feature artifacts are offered. In this subsection, we compare to some of them.

Th\"um et al., in \cite{thum_proof_2011}, presents a composition approach where partial proofs are given in features and then composed to build the complete proofs for an individual product. The specifications are expressed using design by contract \cite{meyer_applying_1992}. The Krakatoa/Why tool \cite{MarchePU04} is used to generate proof obligations that are exported to the Coq proof assistant where proofs are done and verified. 

Our work is in line with the previous approach but it is dedicated to a functional setting whereas Th\"um et al's work involves object-oriented programs. Specification artifacts - expressed as logical formulas - found in a GFML feature module are close to contracts. For example, a refining property 
is close to a refining contract since it includes a former specification and may add a new premise and a new conclusion. However in our work, each proof is complete. It is extended in a new proof in the result of a composition and thus in the resulting product variant.

Another composition framework FEATUREHOUSE is  offered in \cite{apel_featurehouse:_2009}. 
Using the FST (feature structure tree) model, existing artifacts can be composed to construct a new program. The artifacts must have tree structures. In contrast, we presented a framework that allows us to construct products from the artifacts of the features selected by the user. We are interested in expressing feature compositions as  algebraic expressions using composition operations for each kind of artifacts instead of relying on the tree structure. Similar ideas were mentioned in \cite{batory_scaling_2003} \cite{batory_feature_2011}. However, these approaches only focus on constructing products but do not mention how to verify them.

Recent researches \cite{delaware_product_2011}  \cite{delaware_modular_2013}  had proposed some advances in feature composition in the context of meta-theory. However, these tools are dedicated to a very specific domain. Similar to Thum's work, the products in these approaches are verified in Coq. In contrast, implementing feature modules independently of any concrete target language is the purpose of our work. By proposing a generic formal language (GFML) for feature module, the artifacts are easy to reuse and synthesize but don't belong to any concrete language.

\section{Conclusion}\label{sec:conclusion}
In this paper we have described a first step towards the production of a methodology allowing for the development of correct-by-construction product variants according to a FOP paradigm. The contribution is here the definition of the language GFML allowing the developer to write feature modules containing their specifications, code and correctness proofs.
 
These modules are translated to \focalize{} for verification and also for obtaining OCaml operational code. Consequently the product variants we are aiming at will be implemented in OCaml. 
GFML is here introduced to help the developer when he is writing the different feature modules, allowing him to describe the artifacts of a feature with respect to the artifacts of its parents. Realizing this directly in \focalize{} is a difficult task as it is exemplified by the translation scheme presented in a previous section. For the moment, nothing is implemented, all the examples found in this paper (all the possible configurations of the case study) have been obtained by a manual translation and composition.

Next step is to provide an implementation for the GFML Pre-compiler that will automatically translate GFML to \focalize. Then composition of feature modules  will be formally defined and  implemented in the GFML Combiner. 

Another important perspective is to make our methodology independent of the target languages. Regarding this point, an intermediate step could be to adapt our methodology and tools to B or EventB where inheritance is not available. 

\bibliographystyle{eptcs}
\bibliography{fmsple}
\end{document}